\RequirePackage{ifpdf}
\ifpdf 
\documentclass[pdftex]{sigma}
\else
\documentclass{sigma}
\fi

\def\d{{\rm d}}
\def\Re{{\rm Re}\,}                     
                    
\def\ifrac#1#2{{#1/#2}}
\def\m{\phantom{-}}

\numberwithin{equation}{section}

\begin{document}
\allowdisplaybreaks

\renewcommand{\PaperNumber}{068}

\FirstPageHeading

\ShortArticleName{Painlev\'e Analysis and Similarity Reductions for the Magma Equation}

\ArticleName{Painlev\'e Analysis and Similarity Reductions\\ for the Magma Equation}

\Author{Shirley E.~HARRIS~$^\dag$  and Peter A.~CLARKSON~$^\ddag$}
\AuthorNameForHeading{S.E.~Harris  and P.A.~Clarkson}

\Address{$^\dag$~Mathematical Institute, University of Oxford, 24--29 St. Giles',
Oxford, OX1 3LB, UK}
\EmailD{\href{mailto:harris@maths.ox.ac.uk}{harris@maths.ox.ac.uk}}
\Address{$^\ddag$~Institute of Mathematics, Statistics and Actuarial Science, 
University of Kent,\\
$\phantom{^\ddag}$~Canterbury, CT2 7NF, UK}
\EmailD{\href{mailto:P.A.Clarkson@kent.ac.uk}{P.A.Clarkson@kent.ac.uk}}

\ArticleDates{Received September 27,
2006; Published online October 05, 2006}

\Abstract{In this paper, we examine a generalized magma equation for rational  
values of two parameters, $m$ and $n$. Firstly, the similarity reductions are 
found using the Lie group method of inf\/initesimal transformations. The Painlev\'e ODE test
is then applied to the travelling wave reduction, and the pairs of $m$ and $n$ which pass the
test are identif\/ied. These particular pairs are further subjected to the ODE test on their other 
symmetry reductions. Only two cases remain which pass the ODE test for all such symmetry reductions
and these are completely integrable. The case when $m=0$, $n=-1$ is related to the Hirota--Satsuma 
equation and for $m=\tfrac12$, $n=-\tfrac12$, it is a real, generalized,
pumped Maxwell--Bloch equation.}


\Classification{35C05; 35Q58; 37K10} 

\section{Introduction}
Scott and Stevenson \cite{refSS} examine the process of melt migration where
buoyant magma rises through the Earth's mantle. The system is
treated as a two-phase f\/low, with the porous rock matrix able to deform
by creep as the melt rises. The model can be written in terms of the porosity
or liquid volume fraction $\phi$, i.e.\ volume of melt in unit volume of the
f\/luid-rock mixture. The constitutive relations $k=k_0 \phi^n$ and $\eta=\eta_0
\phi^{-m}$ are assumed, where $k$ is the matrix per\-meability, $\eta$ is an
ef\/fective viscosity for the matrix and $k_0$ and $\eta_0$ are constants. After
rescaling~$\phi$ with its uniform background level, so that the background
now corresponds to $\phi=1$,
Scott and Stevenson \cite{refSS} derive the partial
dif\/ferential equation
\begin{gather}
\Delta\equiv\phi_t + (\phi^n+m\phi^{n-m-1}\phi_t\phi_x-\phi^{n-m}\phi_{tx})_x=0, \label{magmaIi}
\end{gather}
and suggest that physically appropriate ranges for $m$ and $n$ are
\begin{gather*}
0\le m\le 1, \qquad 2\le n \le 5.
\end{gather*}
The same equation has also been derived independently for $m=0$ by McKenzie
\cite{refMcKenzie}.

Equation (\ref{magmaIi}) is well known to have solitary wave solutions and
various authors have exami\-ned these.
Nakayama and Mason \cite{refNM92} looked for rarefactive solitary waves
for which $\phi\ge 1$ everywhere, by using a travelling wave reduction.
This leads to
\begin{gather*}
c\left(\d\psi\over\d z\right)^2=f(\psi),
\end{gather*}
for some function $f$ where $z=x-ct$ and $\psi(z)=\phi(x,t)$.
They then applied the boundary conditions
\begin{subequations}\label{magmaIiv}
\begin{gather}
f(1)=0,\label{magmaIiv-a}\\
{\d f\over\d \psi}(1)=0,\label{magmaIiv-b}\\
f(\Psi)=0, \quad \Psi>1,\label{magmaIiv-c}
\end{gather}
and proved that such waves exist if the
index $n>1$ and do not
exist if $0\le n\le 1$, with their analysis restricted to $n\ge 0$.
They go on to derive some solutions
for particular values of pairs $(m,n)$ and to look at the large amplitude approximation. In contrast,
in a later paper~\cite{refNM99}, a~perturbation solution for small amplitude rarefactive
waves is examined and at leading order has the sech-squared form of a single-soliton solution of the
Korteweg--de Vries equation.
Nakayama and Mason \cite{refNM91,refNM94} also
look at compressive solitary waves, for which $0\le \phi\le 1$.
For boundary conditions, they take the
f\/irst two to be the same as those for rarefactive waves (\ref{magmaIiv-a}), (\ref{magmaIiv-b}), but in
place of the last, they instead use
\begin{gather}
{\d^2 f\over\d \psi^2}(1)=0. \label{magmaIiv-d}
\end{gather}
\end{subequations}
Zabusky \cite{refZ} and Jef\/frey and Kakutani \cite{refJK} used this condition for
examining algebraic compressive solitary wave solutions of the modif\/ied
Korteweg--de Vries equation. However, they
also found exponential solitary waves and it therefore appears that this
choice of boundary condition is limiting. Indeed, in this paper we exhibit some new
exponential solitary wave solutions.

In Nakayama and Mason \cite{refNM91}, with the boundary conditions (\ref{magmaIiv-a}), (\ref{magmaIiv-b}), 
(\ref{magmaIiv-d}),
the wave speed~$c$ must be equal to the index $n$. They derive three compressive
solitary wave solutions which tend algebraically to the background $\psi=1$ on
either side and reach $\psi=0$ at their maximum amplitude. The wave with $n=m=2$
has a monotone behaviour whereas the one with $n=m=\tfrac32$ is oscillatory. The
solution for $n=3$ and $m=0$ is not dif\/ferentiable at the point of maximum
amplitude, where there is a cusp. In \cite{refNM94} they examine the necessary
conditions for these compressive solitary waves to exist with the same boundary
conditions as before. They f\/ind that it is necessary to have $n=m>1$ and
plot graphs of the curves for integer values $2$, $3$, $4$ and $5$ for $m$ 
and~$n$. It is found that
the shape does not greatly change, but that the width increases slowly as $n=m$
increases. The half-integer values $\tfrac32$, $\tfrac52$, $\tfrac72$, and $\tfrac92$
are also examined and all prove to have similar oscillatory structure.

Takahashi and Satsuma \cite{refTS} make a change of variables and f\/ind a periodic
wave solution for the choice $n=3$, $m=0$ in terms of elliptic integrals.
They also investigate weak solutions and demonstrate the possibility of both a
hump solution and two travelling wave solutions between dif\/ferent levels with a
sharp wavefront. They note that if the if the $\phi_{tx}$ term in (\ref{magmaIi}) is replaced
by $\phi_{xx}$, it is possible to apply a
transformation to yield the Korteweg--de Vries (KdV) equation. However, the
transformation is not necessarily single-valued and the signif\/icance of this
result is unclear. Takahashi, Sachs and Satsuma \cite{refTSS} show that the travelling wave solutions with
compact support suggested in \cite{refTS} are not physically
feasible, due to singularities in the stress. They use them, however, as
suitable initial conditions for various numerical simulations and demonstrate
the break-up into a series of solitary waves.

Marchant and Smyth \cite{refMS} also examine periodic wave solutions for $n=3$, $m=0$
and develop a modulation theory for slowly varying travelling waves. It is found that
full or partial undular bores are possible and there is good comparison between 
approximate wave envelopes and numerical solutions.

Experiments by Scott, Stevenson and Whitehead \cite{refSSW}, Olson and Christensen \cite{refOC}
and Whitehead \cite{refW} show solitary waves on conduits of buoyant
f\/luid in a more viscous f\/luid. The relevant equation is identical
to the magma migration equations for $m=1$ and $n=2$ and the experimental
results are summarized in Whitehead and Helfrich  \cite{refWH90}. They show
Korteweg--de Vries `soliton-like' collisions (cf.~\cite{refDJ}),
in which two solitary waves interact and then emerge unaltered apart
from a phase-shift. The experiments also imply that an arbitrary initial
condition quickly breaks down into an ordered sequence of solitary waves. These
results raise the question about whether the partial dif\/ferential
equation (\ref{magmaIi}) is completely integrable (i.e.~solvable by the inverse
scattering transform) for any of the parameters~$m$ and~$n$. This is further
reinforced by Whitehead and Helfrich ~\cite{refWH86} who demonstrate that (\ref{magmaIi})
reduces to the KdV equation for small disturbances to the
uniform state.

Barcilon and Richter \cite{refBR} have examined rarefactive waves for $m=0$ and
$n=3$ both analytically and numerically. They concluded that this case did
not have soliton solutions as there was an interaction left behind by the
collision of two solitary waves and they could only f\/ind two conservation
laws. A completely integrable partial dif\/ferential equation such as KdV has an inf\/inite number of
such laws.

Harris  \cite{refHarris} also examines conservation laws for (\ref{magmaIi}), without limiting
the values of $m$ and $n$ to the physically relevant range, and proved that
there were at least three laws for $m=n+1$, $n\neq 0$; at least two laws for
$m=1$, $n\neq 0$ and precisely two laws for all other combinations of~$m$
and $n$. Thus the special cases $m=n+1$  and $m=1$ (with $n\neq 0$ for both
situations) are the only ones which could possibly lead to completely
integrable equations.

In this paper, we apply Painlev\'e analysis to look for any such completely
integrable partial dif\/ferential equations, omitting only the case $m=n=0$ for which the equation is
linear. We start by using the Lie group method of inf\/initesimal transformations to f\/ind
all the similarity reductions of the partial dif\/ferential equation to ordinary dif\/ferential equations.
We can then apply the Painlev\'e ODE test \cite{refARS} to these and they must all have the Painlev\'e property (no movable
singular points apart from poles) for the partial dif\/ferential equation to be completely integrable.

Firstly, we examine the travelling wave reduction. Applying the Painlev\'e ODE test, 
we f\/ind that there are f\/ive rational pairs $(m,n)$ which pass for any value of the 
wave velocity $c$. In addition, there are also four cases which 
require $c=1$, but these can be discounted as candidates for complete integrability.

The f\/ive remaining cases are then subjected to the Painlev\'e ODE test on their 
other similarity reductions. Three of these fail the test at this stage, but the other two
pass the procedure. These last two cases are then checked with the Painlev\'e PDE test 
\cite{refWTC} and are found to be
completely integrable. 

One case, $m=0$, $n=-1$ is related to the Hirota--Satsuma equation (or Shallow Water
Wave I Equation) examined by Clarkson and Mansf\/ield \cite{refCM}.
The second pair, $m=\tfrac12$, $n=-\tfrac12$, yields a~partial dif\/ferential equation 
which can be transformed to 
a special case of the real, generalized, pumped
Maxwell--Bloch equation. This has been previously investigated by Clarkson, Mansf\/ield 
and Milne \cite{refCMM}. 
We note that both pairs satisfy the criterion of $m=n+1$ given by
Harris  \cite{refHarris} for the possibility of more than three conservation laws.

\section{Symmetry analysis}\label{sec2}
The classical method for f\/inding symmetry reductions of partial dif\/ferential
equations is the Lie group method of inf\/initesimal transformations. As this
method is entirely algorithmic, though often both tedious and virtually
unmanageable manually, symbolic manipulation programs have been developed
to aid
the calculations. An excellent survey of the dif\/ferent packages available and a
description of their strengths and applications is  given by Hereman
\cite{refHereman1,refHereman2}. In this paper we use the  {\rm MACSYMA} package {\tt
symmgrp.max} \cite{refCHW} to calculate the
determining equations.

To apply the classical method to equation (\ref{magmaIi}), we consider the
one-parameter Lie group of inf\/initesimal  transformations in ($x,t,\phi$) given by
\begin{gather}
x^* = x+\varepsilon \xi (x,t,\phi) + \mathcal{O} (\varepsilon^2),  \nonumber \\
t^* = t+\varepsilon \tau (x,t,\phi) + \mathcal{O}  (\varepsilon^2), \label{trans}\\
\phi^* =  \phi+\varepsilon \eta (x,t,\phi) + \mathcal{O}  (\varepsilon^2),\nonumber
\end{gather}
where $\varepsilon$ is the group parameter. Then one requires that
this transformation leaves invariant the set
\begin{gather*}
S_{\Delta} \equiv \{ \phi(x,t) : \Delta =0 \} 
\end{gather*}
of solutions of (\ref{magmaIi}). This
yields an overdetermined, linear system of equations for the
inf\/inite\-si\-mals $\xi (x,t,\phi)$, $\tau (x,t,\phi)$ and $\eta (x,t,\phi)$. The
associated Lie algebra is realised by vector f\/ields of the form
\begin{gather}
\mathbf{v} = \xi (x,t,\phi) {\partial\over\partial x} +\tau (x,t,\phi)
{\partial\over\partial t} + \eta (x,t,\phi) {\partial\over\partial \phi}.
\label{vecfi}
\end{gather} 
Having determined the inf\/initesimals, the symmetry variables are
found by solving the characteristic equation
\begin{gather*}
{\d x \over \xi (x,t,\phi)} = {\d t \over \tau (x,t,\phi)}=
 {\d \phi \over \eta (x,t,\phi)}, 
\end{gather*}
which is equivalent to solving the invariant surface condition
\begin{gather*}
\psi \equiv \xi (x,t,\phi) \phi_x+\tau (x,t,\phi) \phi_t -
 \eta (x,t,\phi) =0. 
\end{gather*}
The set $S_{\Delta}$ is invariant under the transformation (\ref{trans}) provided
that $ {\rm pr}^{(3)} \mathbf{v}(\Delta) |_{\Delta \equiv 0} =0$
where ${\rm pr}^{(3)}\mathbf{v}$ is the third prolongation of the vector f\/ield
(\ref{vecfi}), which is given explicitly in terms of~$\xi$,~$\tau$ and~$\eta$ (cf.~\cite{refOlver}). 
This procedure yields a system of 28 determining equations, a
linear homogeneous system of equations, for the inf\/initesimals $\xi(x,t,\phi)$,
$\tau(x,t,\phi)$ and~$\eta(x,t,\phi)$ as given in Table~\ref{infin}; these were
calculated using the MACSYMA package {\tt symmgrp.max} \cite{refCHW}.

\begin{table}[t]\centering
\caption{\label{infin}Inf\/initesimals for the magma equation (\ref{magmaIi}).}
\vspace{1mm}
\begin{tabular}{@{\quad}l@{\quad}l@{\quad}l@{\quad}l@{\quad}}
\hline 
\multicolumn{1}{c}{$m, \  n$}& \multicolumn{1}{c}{$\xi$} & \multicolumn{1}{c}{$\tau$} & \multicolumn{1}{c}{$\eta$} \\
\hline
&&&\\[-3mm]
$n\not=m,\  n\not=0$ & $\alpha_1x+\alpha_0$ & $ -{(n+m-2)\alpha_1\over n-m}t
+\alpha_2 $& ${2\alpha_1 \phi\over n-m}$ \\ [5pt]
$n=m\not=0$  & $\alpha_0$ & $(1-m)\alpha_1t+\alpha_2 $& $\alpha_1\phi $\\[5pt]
$m\not=\tfrac43, n=0 $& $\alpha_1x+\alpha_0$ &$g(t)$ &
$-\,{2\alpha_1 \phi/m}$ \\[5pt]
$m= \tfrac43, n=0 $& $\alpha_2^2x^2+\alpha_1x+\alpha_0 $& $g(t)$ &
$-\tfrac32(2\alpha_2 x+\alpha_1)\phi $\\[5pt]
\hline
\end{tabular}
\end{table}

\section{Painlev\'e ODE test on travelling waves}
In this section, we
use the travelling wave ansatz $\phi(x,t)=\psi(z)$, where
$z=x-ct$ to reduce equation (\ref{magmaIi}) to the ordinary dif\/ferential equation
\begin{gather}
c\psi^2\psi'''+c(n-3m)\psi\psi'\psi''-cm(n-m-1)(\psi')^3 +n\psi^{m+1}\psi'-c\psi^{m-n+2}\psi'=0,\label{magmaIIi}
\end{gather}
where $(\cdot)'$ denotes dif\/ferentiation with respect to $z$. We then 
apply the Painlev\'e ODE test as described in \cite{refARS}, but in order to do this,
we must rewrite (\ref{magmaIIi}) in the form
\begin{gather*}
\psi'''=F(z,\psi,\psi',\psi''), 
\end{gather*}
with the function $F$ analytic in $z$ and rational in its other arguments.
If necessary, a transformation must be made before applying
the test.

\subsection{Integer cases}
We will f\/irst start by restricting our investigation to
$m$ and $n$ being integers so that the criteria on $F$ are satisf\/ied
automatically.
We now follow \cite{refARS} and substitute the ansatz
\begin{gather}
\psi\sim\alpha(z-z_0)^p,\qquad \alpha\neq 0, \label{magmaIIiv}
\end{gather}
into (\ref{magmaIIi}), with $z_0$ being an arbitrary constant.
This gives
\begin{gather}
c\alpha^3p[p(1-m)-1][p(1-m+n)-2](z-z_0)^{3p-3}\nonumber\\
\qquad{}+np\alpha^{m+2}(z-z_0)^{p(m+2)-1}-cp\alpha^{m-n+3}(z-z_0)^{p(m-n+3)-1}=0. \label{magmaIIv}
\end{gather}
It is initially assumed that this is valid in the neighbourhood of a
movable singularity and so
$\Re(p)<0$.
We examine (\ref{magmaIIv}) for all possible dominant balances and
for each case, we substitute
\begin{gather}
\psi\sim \alpha(z-z_0)^p+\beta (z-z_0)^{p+r}, \label{magmaIIvi}
\end{gather}
into the simplif\/ied equation obtained by retaining only the 
dominant terms. Keeping just the linear terms in $\beta$ leads to a cubic
equation for $r$ and the solutions of this are known as the resonances of the 
system.
If $p$ is not an integer and if the ansatz (\ref{magmaIIiv}) is
asymptotic as $z\to z_0$, then the leading order behaviour is that of a
branch point. Similarly, if the resonances are not integers, this also 
indicates the presence of a branch point. If either of these situations occur, 
then the equation is 
not of Painlev\'e type and the case can be rejected.
The details of this procedure are given in Table \ref{pless}, listing $p$,
$\alpha$ and the resonances (apart from $r=-1$, which always occurs), along with 
the conditions on $m$ and $n$ for the balance to be dominant.

\begin{table}[t]\centering
\caption{\label{pless}Results of the Painlev\'e test for $\Re(p)<0$.}
\vspace{1mm}
\begin{tabular}{@{\quad}l@{\quad}l@{\quad}l@{\quad}l@{\quad}}
\hline
\multicolumn{1}{c}{$m, \ n$} & \multicolumn{1}{c}{$p$} & \multicolumn{1}{c}{$\alpha$} & \multicolumn{1}{c}{$r$} \\
\hline
&&&\\[-3mm]
$m>1,\  m+n>2$ &${1\over{1-m}}$&arbitrary& $
 0,\  {{m-1+n}\over{m-1}}$\\[6pt]  $m>1,\  n>1 $&${2\over {1-m}}$&
$\left[{2c\over{m-1}}\right]^{1/(m-1)}$
& $ -2,\ {2n\over{m-1}}$\\[6pt]  $m>n,\  n<1,\  m+n\neq 2$ &${-2\over {m-n}}$
&$\left[{2(2-m-n)\over{(m-n)^2}}\right]^{1/(m-n)}$
&${2\over{m-n}},\ {2(2-n-m)\over{m-n}}$\\[6pt]  
$m>1,\  n=c=1 $&${1\over {1-m}}$&arbitrary& 
$0,\ {m\over{m-1}}$\\[6pt]  
$m>2,\  n=c=1 $&${2\over {2-m}}$&arbitrary&  
$ 0,\ {-2m\over{m-2}}$\\[6pt]  
$m>1,\  n=1,\   c\neq 1$& ${2\over {1-m}}$
&$\left[{2c\over(m-1)(1-c)}\right]^{1/(m-1)}$
&$-2,\ {2\over{m-1}}$\\[6pt]  \hline
\end{tabular}
\end{table}

We now re-examine (\ref{magmaIIv}) but considering
$\Re(p)>0$ so that $\psi$ itself is f\/inite at $z=z_0$ but one of its derivatives
may have a singularity. The results for the dominant balances and corresponding
values of $p$ and $\alpha$ are presented in Table \ref{pmore}.

\begin{table}[t]\centering
\caption{\label{pmore}Results of the Painlev\'e test for $\Re(p)>0$.}
\vspace{1mm}
\begin{tabular}{@{\quad}l@{\quad}l@{\quad}l@{\quad}l@{\quad}}
\hline
\multicolumn{1}{c}{$m, \  n $} & \multicolumn{1}{c}{$p$} & \multicolumn{1}{c}{$\alpha$} & \multicolumn{1}{c}{$r$} \\
\hline
&&&\\[-3mm]
 $m<1,\ m+n<2 $&${1\over{1-m}}$&{\hbox {arbitrary}}& 
 $0,\ {{1-m-n}\over{1-m}}$\\[6pt] 
$1+n-m>0,\  n\ge 0 $&${2\over {1+n-m}}$&{\hbox {arbitrary}}
& $0,\ {2(m+n-1)\over{1+n-m}}$\\[6pt] 
$m<1,\  n<1,\  n\neq 0$&${2\over {1-m}}$
&$\left[{-2c\over{1-m}}\right]^{1/(m-1)}$
&$ -2,\ {-2n\over{1-m}}$\\[6pt] 
$m<n,\  n>1,\  m+n\neq 2 $&${2\over {n-m}}$
&$\left[{2(2-m-n)\over{(m-n)^2}}\right]^{1/(m-n)}$& 
${-2\over{n-m}},\ {2(n+m-2)\over{n-m}}$\\[6pt] 
$m<0,\  n=0 $&${-2\over {m}}$
&$\left[{2(2-m)\over{m^2}}\right]^{1/m}$&
$  {2\over{m}},\ {2(2-m)\over{m}}$\\[6pt] 
$m<1,\  n=c=1$ &$ {1\over {1-m}}$&{\hbox {arbitrary}}& 
$ 0,\ {-m\over{1-m}}$\\[6pt] 
$m<2,\  n=c=1$&$  {2\over {2-m}}$&{\hbox {arbitrary}}&
$ 0,\ {-2m\over{m-2}}$\\[6pt] 
$m<1,\  n=1,\  c\neq 1$& $ {2\over {1-m}}$
&$ \left[{2c\over(1-m)(1-c)}\right]^{1/(m-1)}$&
$ -2,\ {-2\over{1-m}}$\\[6pt] 
\hline
\end{tabular}
\end{table}

Considering the results in Tables \ref{pless} and \ref{pmore}, we
f\/ind that there are ten possibilities for
integer solutions $m$ and $n$ which give rise to dominant balances with 
only integer values of $p$ and these are listed in Table \ref{integer}. 

\begin{table}[t]\centering
\caption{\label{integer}Integer solutions of the Painlev\'e test.}
\vspace{1mm}
\begin{tabular}{@{\quad}l@{\quad}l@{\quad}l@{\quad}l@{\quad}l@{\quad}l@{\quad}}
\hline
\multicolumn{1}{c}{$m $}& \multicolumn{1}{c}{$n$}& \multicolumn{1}{c}{$p$} & 
\multicolumn{1}{c}{$\alpha$} & \multicolumn{2}{c}{$r$} \\
\hline
&&&&\\[-3mm]
0 & $-2$ &$ -1$  &$\pm \sqrt{2}$      &$\m 1$,  &$\m 4$\\
  &      &$\m 1$&{\hbox {arbitrary}}  &$\m 0$,  &$\m 3$\\
  &      &$\m 2$&$-1/2c$                &$-2$,    &$\m 4$\\[6pt] 
0 & $-1$ &$-2$  &6                    &$\m 2$,  &$\m 6$\\
  &      &$\m 1$&{\hbox {arbitrary}}  &$\m 0$,  &$\m 2$\\
  &      &$\m 2$&$-1/2c$              &$-2$,    &$\m 2$\\[6pt] 
0 &$\m 1$&$\m 1$&{\hbox {arbitrary}}  &$\m 0$,  &$\m 0$\\
  &      &$\m 2$&$(c-1)/2c,\, c\neq 1$&$-2$,    &$-2$\\[6pt] 
1 &$-1$  &$-1$  &$\pm 1$              &$\m 1$,  &$\m 2$\\[6pt] 
1 &$\m 0$&$-2$  &2                    &$\m 2$,  &$\m 2$\\[6pt] 
1 &$\m 1$&$\m 2$&{\hbox {arbitrary}}  &$\m 0$,  &$\m 2$\\[6pt] 
1 &$\m 2$&$\m 1$&{\hbox {arbitrary}}  &$\m 0$,  &$\m 2$\\
  &      &$\m 2$&$-1/2$                 &$-2$,    &$\m 2$\\[6pt] 
2 &$\m 1$&$-2$  &$2c/(1-c), \, c\neq 1$&$-2$,   &$\m 2$\\
  &      &$-1$  &{\hbox {arbitrary}}  &$\m 0$,  &$\m 2$\\[6pt] 
2 &$\m 2$&$-2$  &$2c$                 &$-2$,    &$\m 4$\\
  &      &$-1$  &{\hbox {arbitrary}}  &$\m 0$,  &$\m 3$\\
  &      &$\m 2$&{\hbox {arbitrary}}  &$\m 0$,  &$\m 6$\\[6pt] 
2 &$\m 3$&$-2$  &$2c$                 &$-2$,    &$\m 6$\\
  &      &$-1$  &{\hbox {arbitrary}}  &$\m 0$,  &$\m 4$\\
  &      &$\m 1$&{\hbox {arbitrary}}  &$\m 0$,     &$\m 4$\\[6pt]
  \hline
\end{tabular}
\end{table}

The resonances in these cases are all integers and they correspond to the points
in the solution expansion at which arbitrary constants can be introduced. The
resonance $r=-1$ arises from the arbitrariness of $z_0$, whilst if $\alpha$ is 
undetermined, then $r=0$ is another resonance. For each pair ($p$, $\alpha$), we
carry out the solution expansion on the full equation in integral
powers of $z-z_0$, from $(z-z_0)^p$
up as far as $(z-z_0)^{p+r_{\max}}$, where $r_{\max}$ is the largest positive 
resonance.
If an inconsistency arises, this suggests that logarithmic terms
are required in the expansion and hence the ordinary dif\/ferential equation 
fails the Painlev\'e test. A
repeated resonance indicates a logarithmic branch point with arbitrary
coef\/f\/icient and this also fails the test \cite{refARS}.
Negative resonances are not well-understood and if they occur for $p<0$, 
then formally
the ordinary dif\/ferential equation fails the test. In a number of the cases, $r=-2$ occurs in 
conjunction with $p=2$, but this then corresponds to a Taylor series beginning
with a constant term i.e.\ $p=0$, which was not considered and for which 
there are no inconsistencies in the expansion. Thus, these particular negative
resonances are spurious and can be ignored.

There are four integer cases which pass the test, namely $m=0$, $n=-1$; 
$m=0$, $n=-2$; $m=n=c=1$ and $m=2$, $n=c=1$. However, the last
two only pass for one specif\/ic value of the wave speed $c$ and therefore will
not be completely integrable. 

\subsection{Rational cases}
When $m$ and $n$ are not integers, we must f\/irst transform the third order
dif\/ferential equation (\ref{magmaIIi}) so that it is in the correct form for applying
the Painlev\'e ODE test. Using the transformation
\begin{gather}
\psi=w^l, \label{rational}
\end{gather}
where $l$ is an appropriately chosen positive integer, we can obtain a
suitable ordinary dif\/ferential equation for $w$, where $w'''$ is a rational function 
of $w$, $w'$ and $w''$.
We note that if $\psi$ is given by~(\ref{magmaIIvi}), then the above transformation
gives the equivalent expansion for $w$ as
\[
w\sim \alpha_1(z-z_0)^{p_1} + \beta_1(z-z_0)^{p_1+r_1},
\]
where
\begin{gather}
p_1={p/l},\qquad r_1=r. \label{magmaIIxx}
\end{gather}
Thus the resonances are left unaltered by this procedure, but the power $p$ is
changed by a factor of $l$. We can now use the dominant balances and resonances
found for the integer values in Tables~\ref{pless} and~\ref{pmore}, 
bearing in mind (\ref{magmaIIxx}).

We also note that there are no balances in the list which are appropriate for
$m+n=2$, with $m\ge \tfrac32$. This suggests that a simple power is not the correct
choice of ansatz and that we need to incorporate logarithmic terms at leading
order. In fact, we have
\[
\psi\sim \alpha(z-z_0)^{1/(1-m)}[{\rm ln}(z-z_0)]^{1/(2(1-m))},
\]
with
\[
\alpha^{2(m-1)}={1\over 2(1-m)}.
\]
Thus this case fails the Painlev\'e test due to the logarithms.

It is now helpful to draw an $mn$-plot to f\/ind which regions of the plane
will fail the Painlev\'e test.
For negative values of $p$, failure occurs where we have negative resonances
and these occur when
\begin{alignat*}{3}
&m>1, \qquad &&n>1, &\\ 
& n<1, \qquad &&m+n>2, &\\
& m>2, \qquad &&n=1. &
\end{alignat*}
For positive $p$, we can eliminate the cases where $p+r<0$, namely
\begin{alignat*}{3}
& n\ge 0,\qquad &&m+n<0, & \\ 
& m<n, \qquad &&n>1,\qquad m+n<1, &\\
&m<0, \qquad &&n=0,&\\
&m<-1, \qquad &&n=1.&
\end{alignat*}

We note that if $p/l$ is to be an integer, then $p$ must be an integer and we
can use this to rule out further regions in the $mn$-plane. Thus we have
failure when $-1<p<0$, which leads to
\begin{gather*}
m-n>2, \qquad n<1, \qquad m+n\neq 2.
\end{gather*}
Similarly for positive values of $p$, we have failure if $0<p<1$, and this 
gives us the two extra conditions
\begin{alignat*}{3}
& m<0, && \qquad m+n<2,&\\
& m-n<-1,&&\qquad n\ge 0.& 
\end{alignat*}

\begin{figure}
\centerline{\includegraphics[width=8cm]{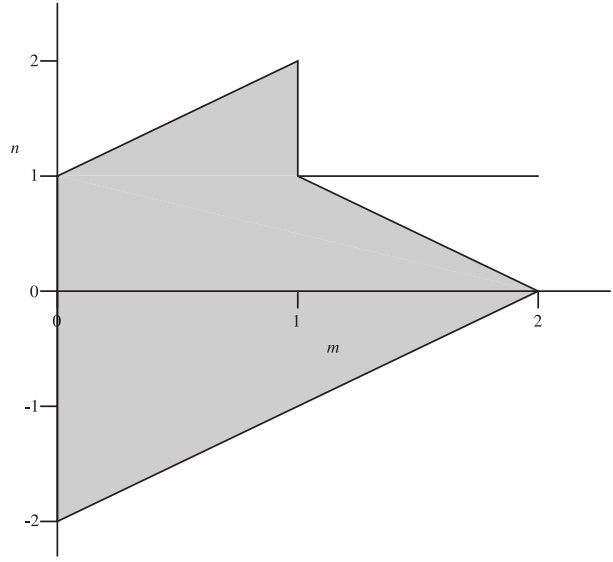}}
\caption{The region of the $mn$-plane which needs to be considered for rational values
of $m$ and $n$.}\label{fig3.1}
\end{figure}

The remaining region is shown in Fig.~\ref{fig3.1} 
and includes the boundaries, the
shaded interior and the line segment $n=1$, $1\le m \le 2$. This can then be 
examined in more detail to f\/ind the rational points which pass
the Painlev\'e test. 
Those which were not examined in the previous subsection
are found to be
\begin{alignat*}{3}
& m={\tfrac12}, \qquad &&n=-{\tfrac12},&\\
&m={\tfrac12}, \qquad && n=0,&\\
&m={\tfrac23}, \qquad &&n=0,& \\
&m={\tfrac32}, \qquad &&n=1=c,& \\
&m={\tfrac43}, \qquad &&n=1=c. &
\end{alignat*}
Again, the last two can be discounted as potential candidates for 
being completely integrable because they only hold for one value of the wave
speed~$c$.

\subsection{Solitary wave solutions}
In looking for solitary wave solutions of the magma equations, it is helpful to
integrate equation~(\ref{magmaIIi}) twice to yield
\begin{gather}
c(\psi')^2 =-{2\,\psi^{1+m}\over{1-m}}+{2c\,\psi^{2+m-n}\over {2-m-n}}
+{2A\,\psi^{1+m-n}\over{1-m-n}}+2B\,\psi^{2m}, \label{magmaIIii}
\end{gather}
for $m\neq 1$, $m+n\neq 2$ and $m+n\neq 1$, where $A$ and $B$ are arbitrary 
constants. Because the scaling
on equation (\ref{magmaIi}) gave a background level of $\psi=1$, we restrict our
interest to this situation. 

Nakayama and Mason \cite{refNM92} examined the existence
of rarefactive solitary waves for which $\psi \ge 1$ and proved that such waves
exist for $n>1$ but do not exist if $0\le n\le 1$. Their proof can easily be 
extended to show that there are no such rarefactive solitary wave solutions for
$n<0$. Nakayama and Mason \cite{refNM94} also looked at the existence of compressive
solitary waves when $0\le \psi\le 1$ and showed that these were only possible
when $m=n>1$. However, their boundary conditions were unnecessarily strict, 
permitting algebraic waves but excluding exponential waves. 

Here, we give some exact solutions which are possible when the 
right hand side of (\ref{magmaIIii}) is a~quartic in $\psi$.  

\begin{description}\itemsep=0pt
\item[(i)] $m=0$, $n=-2$.
\begin{gather*}
\psi=1+{3\sqrt2(c+2)\over{\sqrt{27-(c+5)^2}\,
\cosh\left[\sqrt{{c+2}\over  c}(z-z_0)\right]-\sqrt{2}(1+2 c)}}.
\end{gather*}
This solution describes new exponential compressive waves which exist for $-8\le c<-2$.

\item[(ii)] $m=0$, $n=-1$.
\begin{gather*}
\psi=1-{3(c+1)\over{ c
\cosh\left[\sqrt{{1+ c}\over { c}}(z-z_0)\right]+c }}. 
\end{gather*}
This is also a new compressive solitary wave solution when $-3\le c < -1$ and 
this is illustrated in Fig.~\ref{fig3.2} with $c=-2$.

\begin{figure}[t]
\centering
\includegraphics[width=8cm]{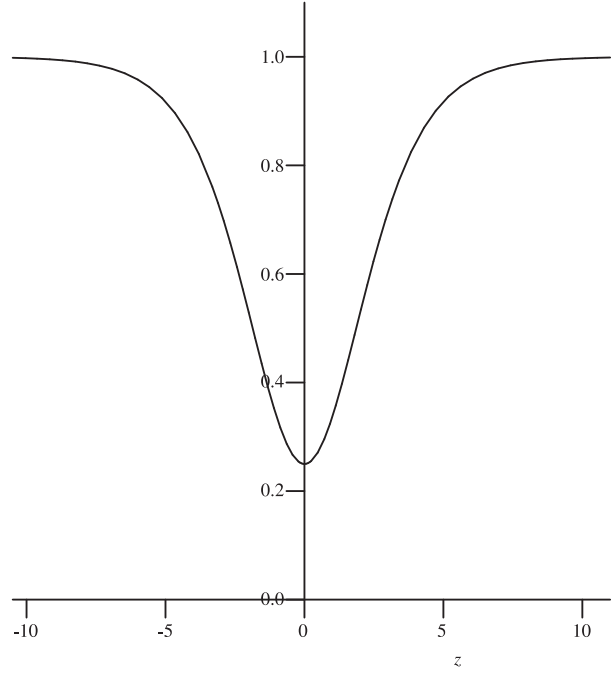}
\caption{A compressive solitary wave for $m=0$, $n=-1$ when $c=-2$.}\label{fig3.2}
\end{figure}

\item[(iii)] $m=2$, $n=2$.
\begin{subequations}\label{magmaIIxxx}
\begin{gather}
\psi=1- {{3(c-2)}\over{ \pm (1-c)\cosh\left[\sqrt{{c-2}\over c}(z-z_0)\right] +
2c-5}}, \label{magmaIIxxx-a}\\
\psi=1-{12\over{12+(z-z_0)^2}}.\label{magmaIIxxx-b} 
\end{gather}
\end{subequations}
Equation (\ref{magmaIIxxx-a}) with the positive sign gives a
rarefactive solitary wave which exists for $2<c<4$ and agrees with that
found by Nakayama and Mason \cite{refNM92}. However, with the minus sign, it represents a new 
exponential compressive
wave solution of amplitude 1 for $c<0$ and $c>2$ and on letting
$c\to 2$, it gives the algebraic wave (\ref{magmaIIxxx-b})
seen in Nakayama and Mason \cite{refNM91}.
These waves are shown in Fig.~\ref{fig3.3}, for dif\/ferent values of the wave speed~$c$.
\end{description}

\begin{figure}
\centering
\includegraphics[width=8cm]{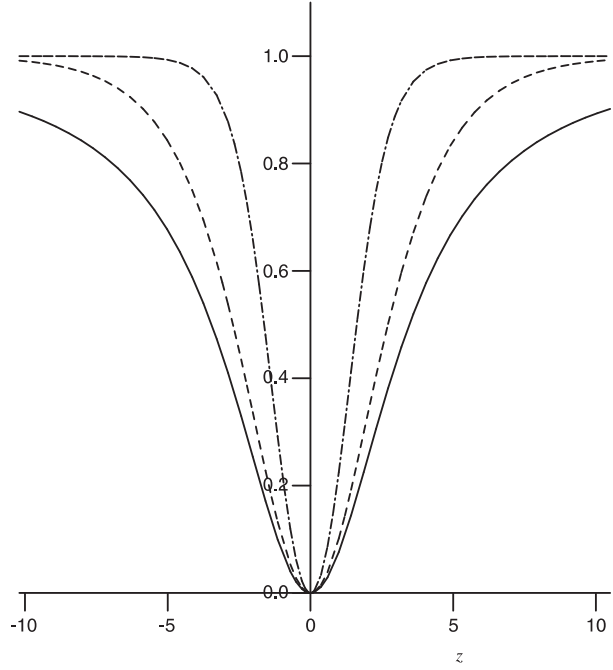}
\caption{ Solitary waves for $m=2$, $n=2$, which are exponential for $c=-2$ 
($-\cdot -$) and $c=3$ ($-\,-\,-$) and algebraic when $c=2$ (------).}\label{fig3.3}
\end{figure}

We now look at other rational values of $m$ and $n$.
By using the substitution (\ref{rational}) to transform equation 
(\ref{magmaIIii}) into an equation for $w$ of the form
\[
\left(w'\right)^2=f(w),
\]
and looking for values of $m$ and $n$ which give $f(w)$ as a quartic 
polynomial, we can f\/ind further solutions of this form.
It turns out that only $l=2$ is possible for real solutions, and the equation 
for $w$ is
\begin{gather*}
\left(w'\right)^2=-{w^{2m}\over{2c(1-m)}}+{w^{2+2m-2n}\over{2(2-m-n)}}
+{Aw^{2m-2n}\over{2c(1-m-n)}}+{B\over{2c}}w^{4m-2}.
\end{gather*}
The values of $m$ for which both the f\/irst and last terms on the right
hand side can give a quartic in $w$ are $m=\tfrac12$ and $m=\tfrac32$ (with $m=1$
being excluded as the denominator of the f\/irst term would vanish). Choosing
appropriate values for $A$ and $B$ so that there is a double root of the
right hand side at $w=1$, we can f\/ind solutions for $w$ and hence $\psi$. 

\begin{description}\itemsep=0pt
\item[(iv)] $m=\tfrac12$, $n=-\tfrac12$, with $c<-\tfrac12$. 
\begin{gather*}
\psi=\left(1+{{1+2c}\over{\sqrt{-\tfrac12c}\,
\cosh\left[{\sqrt{{1+2c}\over{2c}}(z-z_0)}\right]-c}}\right)^2
\end{gather*}
This is a compressive wave which is
oscillatory for $c<-2$ and represents a new solution. (See Fig.~\ref{fig3.4}.)

\begin{figure}[th]
\centering
\includegraphics[width=8cm]{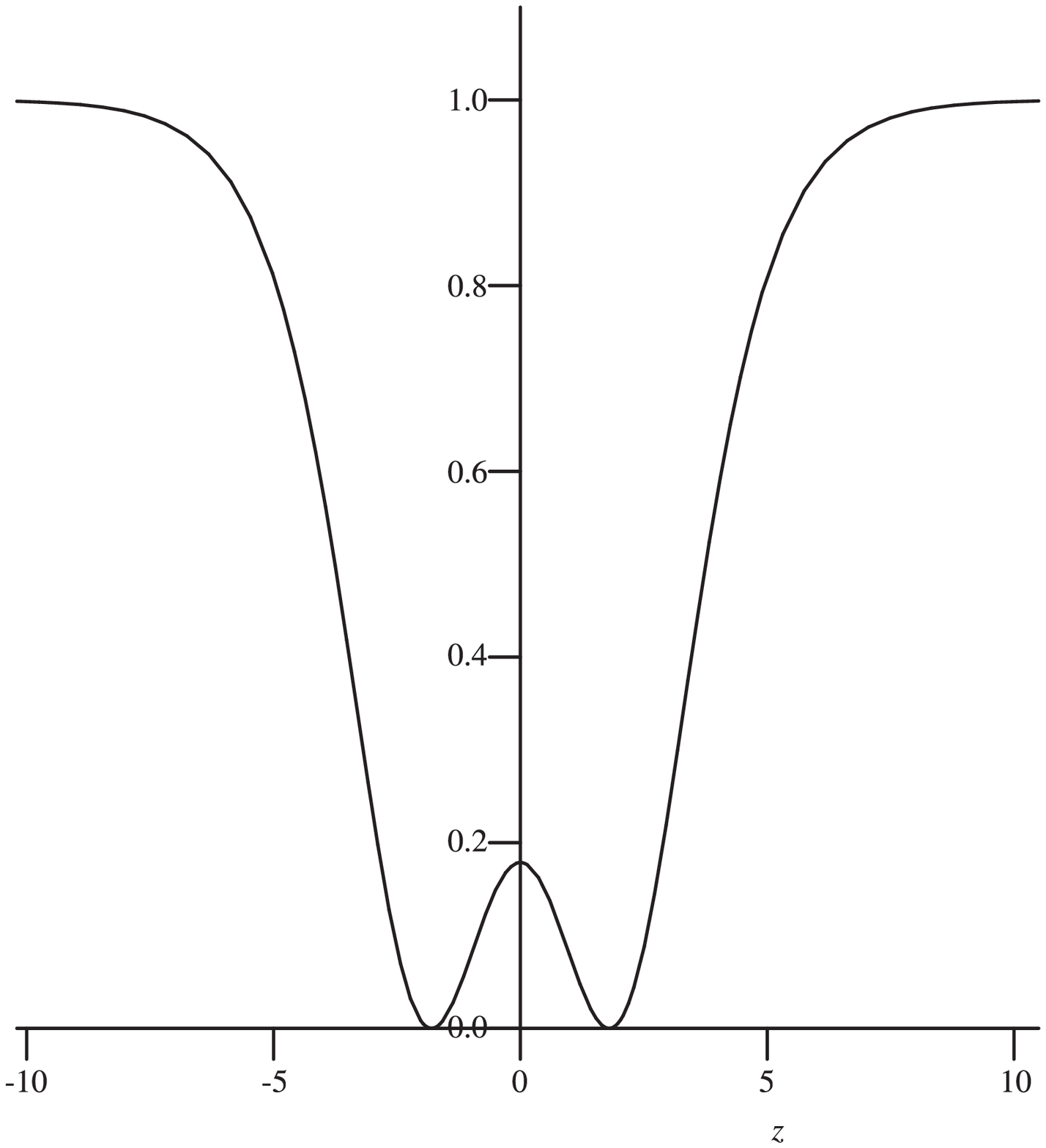}
\caption{An oscillatory solitary wave for $m=-{1\over 2}$, $n=-{1\over 2}$ when $c=-6$.}\label{fig3.4}
\end{figure}

\item[(v)] $m=\tfrac12$, $n=0$.
\begin{gather*}
\psi=\left(1-{6\over{\cosh (z-z_0)+1}}\right)^2. 
\end{gather*}
This new solution is oscillatory in nature, has both compressive and 
rarefactive regions and the shape is independent of velocity. 
(See Fig.~\ref{fig3.5}.)

\begin{figure}[th]
\centering
\includegraphics[width=8cm]{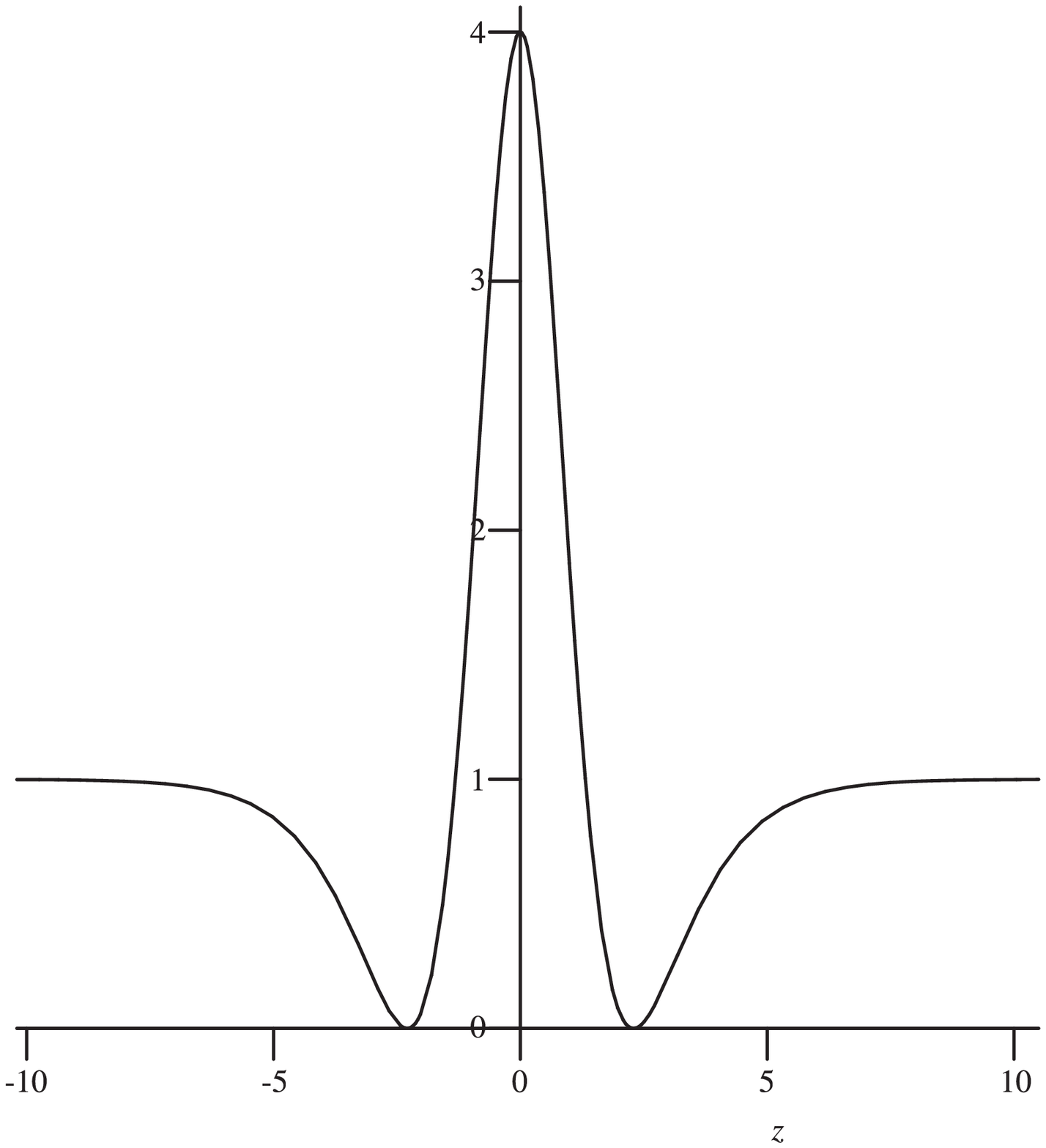}
\caption{A solitary wave with compressive and rarefactive regions for $m={1\over 2}$, $n=0$.}\label{fig3.5}
\end{figure}

\item[(vi)] $m=\tfrac32$, $n=1$.
\begin{gather*}
\psi=\left(1-{6\over{\cosh\left[{\sqrt{{c-1}\over
c}(z-z_0)}\right]+5}}\right)^2. 
\end{gather*}
This is a compressive wave with amplitude 1 for $c<0$ or $c>1$. 

\item[(vii)] $m=\tfrac32$, $n=\tfrac32$.
\begin{subequations}\label{magmaIIxxxvi}
\begin{gather}
\psi=\left(1+{{2c-3}\over{\pm \sqrt{{c-1}\over 2}\cosh\left[{\sqrt{{2c-3}\over
2c} (z-z_0)}\right]+2-c}}\right)^2, \label{magmaIIxxxvi-a}\\
\psi=\left(1-{12\over{(z-z_0)^2+9}}\right)^2.\label{magmaIIxxxvi-b}
\end{gather}
\end{subequations}

Equation (\ref{magmaIIxxxvi-a})
represents rarefactive waves with the positive root
if $3>c>\tfrac32$ (found by Nakayama and Mason \cite{refNM92} 
whereas it gives new compressive oscillatory waves for $c>\tfrac32$ for the negative root.
As $c \to \tfrac32$ for the
compressive wave, we f\/ind the algebraic solution~(\ref{magmaIIxxxvi-b})
which is given in Nakayama and Mason \cite{refNM91}. 

\end{description}

\section{Other scaling reductions}
For the f\/ive cases which pass the Painlev\'e ODE test on the travelling wave reduction, we
now look at the test on the other symmetry reductions found in Section~\ref{sec2}.

\begin{description}\itemsep=0pt
\item[(i)] $m=0$, $n=-2$.
Using the reduction,
\[
z=x t^{\ifrac12}, \qquad \phi(x,t)=t^{\ifrac12} w(z)
\]
leads to the equation
\begin{gather}
zww'''-2zw'w'' - zw^3w' + 3ww'' - 4(w')^2 + 4w' - w^4 =0.\label{eqcasei}
\end{gather}

\item[(ii)] $m=0$, $n=-1$.
Here the appropriate reduction is 
\[z=x t^{\ifrac13}, \qquad \phi(x,t)=t^{\ifrac23} w(z),\]
leading to the equation
\begin{gather}
zww''' -zw'w'' - zw^2w' + 4ww'' - 3(w')^2 + 3w' - 2w^3 =0.\label{eqcaseii}
\end{gather}

\item[(iii)] $m=\tfrac12$, $n=-\tfrac12$.
In this case, we use
\[z=x t^{\ifrac12}, \qquad \phi(x,t)=t w^2(z)\] 
to obtain
\begin{gather}
zww'''-zw'w'' - zw^3w' + 3ww'' - 2(w')^2 + w' - w^4 =0.\label{eqcaseiii}
\end{gather}

\item[(iv)] $m=\tfrac12$, $n=0$.
The correct reduction for this case is
\[z=x h(t), \qquad \phi(x,t)=x^{-4}w^2(z)\]
where $h$ is an arbitrary function and this leads to the equation
\begin{gather}
z^2w'''-2zw''-ww'+2w'=0.\label{eqcaseiv}
\end{gather}

\item[(v)] $m=\tfrac23$, $n=0$.
Lastly, we have the reduction
\[
z=x h(t), \qquad \phi(x,t)=x^{-3}w^3(z)
\]
yielding
\begin{gather}
z^2w'''-w^2w'=0.\label{eqcasev}
\end{gather}
\end{description}

As in the travelling wave reductions, we again use the ansatz
\begin{gather*}
\psi\sim\alpha(z-z_0)^p,\qquad \alpha\neq 0, 
\end{gather*}
to f\/ind the dominant balances for $\Re(p)<0$ and the resulting resonances, with $r=-1$
always occurring due to the arbitrariness of $z_0$. The
results are shown in Table~\ref{tab4.1} and give integer values for $p$ and $r$. 

\begin{table}[t]\centering
\caption{Results of the Painlev\'e test for $\Re(p)<0$.}\label{tab4.1}
\vspace{1mm}
\begin{tabular}{l@{\qquad}l@{\qquad}l@{\qquad}l@{\qquad}l@{\qquad}l} \hline
\multicolumn{1}{c}{$m$} &  \multicolumn{1}{c}{$n$} & \multicolumn{1}{c}{$p$} & 
\multicolumn{1}{c}{$\alpha$} & \multicolumn{2}{c}{$r$}\\
\hline
&&&&&\\[-3mm]
0&  $-2$ &$-1$&$\pm \sqrt{2}$& $1$,&$4$\\[6pt] 
0&$-1$ &$-2$&$6$& $2$,&$6$\\[6pt] 
${1\over2}$&$-{1\over2}$ &$-1$&$\pm 2$&$2$, &$4$\\[6pt] 
${1\over2}$&$0$ &$-2$&$12\, z_0^2$& $4$, &$6$\\[6pt] 
${2\over3}$&$0$ &$-1$&$\pm \sqrt{6}\, z_0$ & $3$, &$4$\\[4pt]
\hline
\end{tabular}
\end{table}

However, for
cases (\ref{eqcasei}), (\ref{eqcaseiv}) and (\ref{eqcasev}), the solution expansion on the full 
equation as far as powers of
$(z-z_0)^{p+r_{\max}}$ gives inconsistencies, indicating that logarithmic terms are required
in the expansion. Therefore, these cases fail the Painlev\'e test. This leads to only
cases (\ref{eqcaseii}) and~(\ref{eqcaseiii}) remaining, which both satisfy the condition $m=n+1$ 
found by Harris~\cite{refHarris}. 

\section{Painlev\'e PDE test}
In this section, we consider the two remaining cases and relate them to known completely integrable
equations.

\begin{description}\itemsep=0pt
\item[(i)] $m=0$, $n=-1$.
A slightly more general equation is considered, namely
\begin{gather}
\phi \phi_{xxt} - \phi_x \phi_{xt} - \phi^2 \phi_t + \mu \phi_x=0, \label{eqcasea}
\end{gather}
which reduces to the magma equation for $m=0$, $n=-1$ upon setting $\mu = 1$.
\end{description}

It is straightforward to apply the Painlev\'e PDE test, as described by Weiss, Tabor and Carnevale 
\cite{refWTC}, to (\ref{eqcasea}) and it is found to pass the test for any values of $\mu$.
We note that (\ref{eqcasea}) can be written in the form
\[
 \left[\left(\phi_{xt} -\mu \right)/\phi\right]_x - \phi_t=0.
\]
Then making the transformation
$\phi=v_x$ and integrating once gives
\begin{gather*}
v_{xxt} - v_xv_t - \mu =0,
\end{gather*}
where we have set the function of integration to zero, without loss of
generality. A simple transformation 
\[
v(x,t)=- u(x,t) -x-t
\]
then leads to the shallow water wave (SWWI) equation
\begin{gather}
u_{xxxt}+ u_x u_{xt} +  u_t u_{xx}-u_{xt} -u_{xx}=0,\label{eqswwi}
\end{gather}
which was considered by Clarkson and Mansf\/ield \cite{refCM}.
This is also equivalent to the Hirota--Satsuma equation
\begin{gather}
\eta_{xxt} + \eta\eta_t - \eta_x\partial_x^{-1}(\eta_t) - \eta_x - \eta_t
=0,\label{eqhs}
\end{gather}
where $\partial^{-1}_x(f)(x) = \int_x^\infty f(y)\,\d y$, which
was discussed by Hirota and Satsuma \cite{refHS}. 

Clarkson and Mansf\/ield
 \cite{refCM} have discussed symmetry reductions of (\ref{eqswwi}) obtainable using the
classical Lie method and the nonclassical method of \cite{refBC}.
They gave a catalogue of classical and nonclassical symmetry  reductions and
exact solutions of (\ref{eqswwi}). Of
particular interest are a plethora of solutions of (\ref{eqswwi}) possessing a rich
variety of qualitative behaviours. These arise as nonclassical symmetry
reductions and all of which look like a two-soliton solution as $t\to-\infty$,
yet are radically dif\/ferent as $t\to\infty$. These results have important
implications with regard to numerical analysis and suggest that solving (\ref{eqswwi})
numerically could pose some fundamental dif\/f\/iculties. An exponentially small
change in the initial data yields a fundamentally dif\/ferent solution as
$t\to\infty$. How can any numerical scheme in current use cope with such
behaviour? (See  \cite{refCM} for further details.)

The Lax pair for the Hirota--Satsuma equation (\ref{eqhs}) is the third
scattering order problem \cite{refConte}
\begin{subequations}\label{scpiab}
\begin{gather}
\psi_{xxx} - \left(1+\tfrac12\eta\right)\psi_x
=\lambda\psi,\label{scpia}\\
3\lambda\psi_t = [1-\partial^{-1}_x(\eta_t)]\psi_{xx} -
\eta_{t}\psi_x.\label{scpib}
\end{gather}
\end{subequations}
We remark that (\ref{scpia}) is similar to the scattering problem
\begin{gather*}
\psi_{xxx} + \tfrac14(1+6u)\psi_x +\tfrac34\left[u_x -
{\rm i}\sqrt{3}\,\partial_x^{-1}(u_t)\right]\psi= \lambda\psi,
\end{gather*}
which is the scattering problem for the Boussinesq equation
\begin{gather*}
u_{xxxx} + 3(u^2)_{xx} + u_{xx} = u_{tt},
\end{gather*}
and has been comprehensively studied by Deift, Tomei and Trubowitz \cite{refDTT}.

\begin{description}\itemsep=0pt
\item[(ii)] $m=\tfrac12$, $n=-\tfrac12$.
Similarly, we consider a slightly more general equation
\begin{gather}
uu_{xxt} - u_x u_{xt} - u^3u_t + \mu u_x=0, \label{eqcaseb}
\end{gather}
where setting $\phi(x,t)=u^2$ and $\mu=\tfrac12$ yields the magma equation for
$m=\tfrac12$, $n=-\tfrac12$.
\end{description}
Again, the Painlev\'e PDE test due to Weiss, Tabor and Carnevale \cite{refWTC} can be applied
and~(\ref{eqcaseb}) is found to pass for any $\mu$.
Clarkson, Mansf\/ield and Milne \cite{refCMM} showed that (\ref{eqcaseb}) arises as a~symmetry
reduction of the $(2+1)$-dimensional Sine-Gordon system
\begin{gather*}
\left({\Theta_{XT} \over \sin\,\Theta}\right)_X
-\left({\Theta_{YT} \over \sin\Theta}\right)_Y
-{\Theta_YR_{XT}-\Theta_XR_{YT}\over \sin^2\Theta}= 0,\\
\left({R_{XT} \over \sin\Theta}\right)_X
-\left({R_{YT} \over \sin\Theta}\right)_Y
-{\Theta_Y\Theta_{XT}-\Theta_X\Theta_{YT}\over \sin^2\Theta}=0,
\end{gather*}
derived by Konopelchenko and Rogers \cite{refKR91,refKR93}. The special case of
(\ref{eqcaseb})
with $\mu\equiv0$ is equivalent to the sine-Gordon equation
\begin{gather}
u_{xt} = \sin u,\label{eqSG}
\end{gather}
which is one of the fundamental soliton equations solvable by inverse
scattering using the AKNS method \cite{refAKNS}.

Clarkson, Mansf\/ield and Milne \cite{refCMM} show that a Lax pair associated with
(\ref{eqcaseb}) is given by
\begin{subequations}\label{eqwlax-ab}
\begin{gather}
{\boldsymbol \psi}_x = \left\{-{\rm i}\lambda{\boldsymbol \sigma}_3 -
\tfrac12{\rm i} u{\boldsymbol \sigma}_2\right\}{\boldsymbol \psi},\label{eqwlax}\\
{\boldsymbol \psi}_t = \left\{{\rm i}{u_{xt}-\mu\over 4\lambda u}{\boldsymbol \sigma}_3  + {{\rm i} u_t\over
4\lambda}{\boldsymbol \sigma}_1 - {{\rm i} \mu\over8\lambda^2}{\boldsymbol \sigma}_2
\right\}{\boldsymbol \psi} =0.\label{eqwlax-b}
\end{gather}
\end{subequations}
where ${\boldsymbol \sigma}_1$, ${\boldsymbol \sigma}_2$ and ${\boldsymbol \sigma}_3$ are the Pauli spin matrices
given by
\[
{\boldsymbol \sigma}_1=\left(\begin{array}{cr} 0 &  1 \\ 1 & 0 \end{array}\right),\qquad
{\boldsymbol \sigma}_2=\left(\begin{array}{cr} 0 &  -{\rm i} \\ {\rm i} & 0 \end{array}\right),\qquad
{\boldsymbol \sigma}_3=\left(\begin{array}{cr} 1 &  0 \\ 0 & -1 \end{array}\right).
\]
Equations \eqref{eqwlax-ab}
are compatible, i.e.\
${\boldsymbol \psi}_{xt}={\boldsymbol \psi}_{tx}$ provided that $u(x,t)$ satisf\/ies (\ref{eqcaseb}). We
remark
that the Lax pair \eqref{eqwlax-ab}
reduces to that for the sine-Gordon equation
(\ref{eqSG}) if $\mu=0$ \cite{refAKNS}. Further we note
that
the spectral problem (\ref{eqwlax}) is the standard AKNS spectral problem whilst
if $\mu\not=0$, then \eqref{eqwlax-b}
involves powers of both $\lambda^{-1}$
and $\lambda^{-2}$.

\section{Discussion}
We have performed a comprehensive Painlev\'e analysis of the generalized magma
equation
\[
\phi_t + (\phi^n+m\phi^{n-m-1}\phi_t\phi_x-\phi^{n-m}\phi_{tx})_x =0, 
\]
and found that there are only two pairs of values of the indices $m$ and $n$ for which the equation 
is completely integrable. The case $m=0$, $n=-1$ is related to the Hirota--Satsuma equation 
and was studied by Clarkson and Mansf\/ield \cite{refCM}. 
The Hirota bilinear form,
multi-soliton solutions and the Lax pair are known for this equation.
The situation when $m=\tfrac12$, $n=-\tfrac12$ is less clear cut. It is related by a
simple transformation to a special case of the real, generalized, pumped 
Maxwell--Bloch system examined by Clarkson, Mansf\/ield and Milne~\cite{refCMM}. It has
an non-isospectral Lax Pair and a single soliton solution, but the Hirota bilinear 
form and multi-soliton solutions are as yet unknown.

\subsection*{Acknowledgements}
{It is a pleasure to thank Elizabeth Mansf\/ield and Andrew Pickering for 
their helpful comments and discussions. SEH gratefully acknowledges the support of a Darby
Fellowship at Lincoln College, Oxford and a Grace Chisholm Young Fellowship from the 
London Mathematical Society for the period in which this research was carried out. PAC thanks
the Isaac Newton Institute, Cambridge for their hospitality during his visit as part of the
programme on ``\emph{Painlev\'e Equations and Monodromy Problems}'' when this paper was
completed.}

\LastPageEnding


\begin{thebibliography}{99}
\footnotesize

\bibitem{refAKNS}
Ablowitz M.J., Kaup D.J., Newell A.C., Segur H.,
The inverse scattering transform -- Fourier analysis for nonlinear problems,
{\it Stud. Appl. Math.}, 1974, V.53, 249--315.

\bibitem{refARS} Ablowitz M.J., Ramani A., Segur H.,
A connection  between nonlinear evolution equations and ordinary dif\/ferential
equations of P-type. I, {\it J. Math. Phys.}, 1980, V.21, 715--721.

\bibitem{refBR} Barcilon V., Richter F.M.,
Nonlinear waves in compacting media,
{\it J. Fluid Mech.}, 1986, V.164, 429--448.


\bibitem{refBC} Bluman G.W., Cole J.D., The general similarity solution of the heat equation,
{\it J. Math. Mech.}, 1969, V.18, 1025--1042.

\bibitem{refCHW} Champagne B., Hereman W., Winternitz P.,
The computer calculation of Lie point symmetries of large systems of dif\/ferential equations,
{\it Comput. Phys. Comm.}, 1991, V.66, 319--340.

\bibitem{refCM} Clarkson P.A., Mansf\/ield E.L., On a shallow water wave equation,
{\it Nonlinearity}, 1994, V.7, 975--1000, 
\href{http://arxiv.org/abs/solv-int/9401003}{\mbox{solv-int/9401003}}.

\bibitem{refCMM} Clarkson P.A., Mansf\/ield E.L., Milne A.E.,
Symmetries and exact solutions of a $2+1$-dimensional Sine-Gordon system, {\it Phil. Trans. R. Soc. Lond.~A},
1996, V.354, 1807--1835, \href{http://arxiv.org/abs/solv-int/9412003}{solv-int/9412003}.

\bibitem{refConte} Conte R., Musette M., Grundland A.M.,
B\"acklund transformation of partial dif\/ferential equations from the
Painlev\'e-Gambier classif\/ication. II. Tzitzeica equation,
{\it J. Math. Phys.}, 1999, V.40, 2092--2106.

\bibitem{refDTT} Deift P., Tomei C., Trubowitz E.,
Inverse scattering and the Boussinesq equation, {\it Commun. Pure Appl. Math.},
1982, V.35, 567--628.

\bibitem{refDJ} Drazin P.G., Johnson R.S., Solitons: an introduction, 
Cambridge, Cambridge University Press, 1989.

\bibitem{refHarris} Harris S.E., Conservation laws for a nonlinear wave equation,
{\it Nonlinearity},  1996, V.9, 187--208.

\bibitem{refHereman1} Hereman W., Symbolic software for Lie symmetry
analysis, in CRC Handbook of Lie Group Analysis of Differential Equations. III. New Trends
in Theoretical Developments an Computational Methods, Editor N.H.~Ibragimov,
 Boca Raton, CRC Press, 1996, Chapter XII, 367--413.

\bibitem{refHereman2} Hereman W., Review of 
symbolic software for Lie symmetry analysis, 
{\it Math. Comput. Modelling}, 1997, V.25, 115--132.

\bibitem{refHS} Hirota R., Satsuma J., $N$-soliton solutions of model equations for shallow water waves,
{\it J. Phys. Soc. Japan},  1976, V.40, 611--612.

\bibitem{refJK} Jef\/frey A., Kakutani T.,
Weak nonlinear dispersive waves: a discussion centered around the Korteweg--de Vries equation,
{\it SIAM Rev.}, 1972, V.14, 582--643.

\bibitem{refKR91} Konopelchenko B.G., Rogers C.,
On $2+1$-dimensional nonlinear systems of Loewner type,
{\it Phys. Lett.~A}, 1991, V.158, 391--397.

\bibitem{refKR93} Konopelchenko B.G., Rogers C.,
On generalized Loewner systems: novel integrable equations in $2+1$-dimensional,
{\it J. Math. Phys.}, 1993, V.34, 214--242.

\bibitem{refMS} Marchant T.R., Smyth N.F., Approximate solutions for magmon propagation from a reservoir,
{\it IMA J. Appl. Math.}, 2005, V.70, 796--813.

\bibitem{refMcKenzie} McKenzie D.P., 
The generation and compaction of partially molten rock,
{\it J. Petrol.}, 1984, V.25, 713--765.

\bibitem{refNM91} Nakayama M., Mason D.P.,
Compressive solitary waves in compacting media,
{\it Internat. J. Non-Linear Mech.}, 1991, V.26, 631--640.
 
\bibitem{refNM92} Nakayama M., Mason D.P.,
Rarefactive solitary waves in two--phase f\/luid f\/low of compacting media,
{\it Wave Motion}, 1992, V.15, 357--392.

\bibitem{refNM94} Nakayama M., Mason D.P., On the existence of compressive solitary waves in compacting media,
{\it J. Phys.~A: Math. Gen.}, 1994, V.27, 4589--4599.

\bibitem{refNM99} Nakayama M., Mason D.P., 
Perturbation solution for small amplitude solitary waves in two-phase f\/luid f\/low of compacting media,
{\it J. Phys. A: Math. Gen.}, 1999, V.32, 6309--6320.

\bibitem{refOC} Olson P., Christensen U.,
Solitary wave propagation in a f\/luid conduit within a viscous matrix,
{\it J. Geophys. Res.}, 1986, V.91, 6367--6374.

\bibitem{refOlver} Olver P.J.,
Applications of Lie groups to dif\/ferential
equations, 2nd ed., {\it Graduate Texts Math.}, Vol.~107, New York, Springer, 1993.

\bibitem{refSS} Scott D.R., Stevenson D.J.,
Magma solitons, {\it Geophys. Res. Lett.},  1984, V.11, 1161--1164.
 
\bibitem{refSSW} Scott D.R., Stevenson D.J., Whitehead J.A.,
Observations of solitary waves in a viscously deformable pipe,
{\it Nature}, 1986, V.319, 759--761.

\bibitem{refTSS} Takahashi D., Sachs J.R., Satsuma J.,
Properties of the magma and modif\/ied magma equations,
{\it J. Phys. Soc. Japan}, 1990, V.59, 1941--1953.
 
\bibitem{refTS} Takahashi D., Satsuma J.,
Explicit solutions of magma equation,
{\it J. Phys. Soc. Japan}, 1988, V.57, 417--421.

\bibitem{refWTC} Weiss J., Tabor M., Carnevale G.,
The Painlev\'e property for partial dif\/ferential equations,
{\it J. Math. Phys.},  1983, V.24, 522--526.

\bibitem{refW} Whitehead J.A.,
A laboratory demonstration of solitons using a vertical watery conduit in syrup,
{\it Amer. J. Phys.},  1987, V.55, 998--1003.

\bibitem{refWH86} Whitehead J.A., Helfrich K.R.,
The Korteweg--de Vries equation from conduit and magma migration equations,
{\it Geophys. Res. Lett.}, 1986, V.13, 545--546.

\bibitem{refWH90} Whitehead J.A., Helfrich K.R., Magma waves and diapiric dynamics,
in Magma Transport and Storage, Editor M.P.~Ryan, Wiley \& Sons, 1990, 53--76.

\bibitem{refZ} Zabusky N.J.,
A synergetic approach to problems of nonlinear dispersive wave propagation and interaction,
in Proc. Symp. Nonlinear Partial Dif\/ferential Equations,
Editor W.~Ames, New York, Academic Press,  1967, 223--258.


\end{thebibliography}
\end{document}